\documentclass[article,reprint,amsmath,amssymb,raggedbottom,letterpaper,final,superscriptaddress,citeautoscript,floatfix,notitlepage]{achemso}
\usepackage{amssymb,amsmath,bm}
\usepackage[usenames,dvipsnames]{xcolor}
\usepackage[bookmarks=false,colorlinks]{hyperref}
  \hypersetup{linkcolor=black,citecolor=black,filecolor=black,urlcolor=black}
\usepackage{graphicx}
\usepackage{booktabs}
\usepackage{tabularx}
\usepackage{etoolbox}
\makeatletter
\patchcmd{\frontmatter@abstract@produce}
  {\vskip200\p@\@plus1fil
  \penalty-200\relax
  \vskip-200\p@\@plus-1fil}
  {}
  {}
  {}
\makeatother

\title{Accelerating Discovery of Metal-Insulator Transition Compounds Using Physics-Informed Machine Learning}

\author{Alexandru B.\ Georgescu}
\email{georgesc@iu.edu}
\affiliation{%
Department of Chemistry, 800 East Kirkwood Avenue, Indiana University, Bloomington, Indiana 47405, United States}
\affiliation{Department of Materials Science and Engineering, Northwestern University, Evanston, IL 60208, USA}
\author{Peiwen Ren}
\affiliation{Department of Materials Science and Engineering, Northwestern University, Evanston, IL 60208, USA}
\author{Harshul Bhatt}
\affiliation{%
Department of Chemistry, 800 East Kirkwood Avenue, Indiana University, Bloomington, Indiana 47405, United States}
\author{Christopher Karpovich}
\affiliation{Department of Materials Science and Engineering, Massachusetts Institute of Technology, Cambridge, MA 02139 USA}
\author{Bipasa Samanta}
\affiliation{%
Department of Chemistry, 800 East Kirkwood Avenue, Indiana University, Bloomington, Indiana 47405, United States}
\author{Elsa Olivetti}
\affiliation{Department of Materials Science and Engineering, Massachusetts Institute of Technology, Cambridge, MA 02139 USA}
\author{James M.\ Rondinelli}
\email{jrondinelli@northwestern.edu}
\affiliation{Department of Materials Science and Engineering, Northwestern University, Evanston, IL 60208, USA}

\begin{document}
%\preprint{}


%\thanks{}

\newcommand{\pr}[1]{{\color{Orange}[\textbf{PR: }\textit{{#1}}]}}
\newcommand{\jmr}[1]{\textcolor{blue}{{\sffamily #1}}}
\newcommand{\abg}[1]{{\color{Purple}[\textbf{ABG: }\textit{{#1}}]}}
\newcommand{\cmo}[1]{{CaMn$_2$O$_4$}}
\newcommand{\cfo}[1]{{Ca$_2$Fe$_3$O$_8$}}
\newcommand{\cco}[1]{{CaCo$_2$O$_4$}}
\newcommand{\lco}[1]{{LaCoO$_3$}}

%\date{\today}

\begin{abstract}
Metal-insulator transition (MIT) materials are a useful platform for emerging microelectronic, optoelectronic, and neuromorphic devices, but their discovery is hindered by the high computational cost of electronic structure modeling, the complexity of underlying mechanisms, and the challenges of experimental validation. Here, we present a physics-informed machine learning framework that accelerates the discovery of thermally driven MIT materials. Using a trained classifier, we screen
a crystal structure database to identify promising candidates for higher fidelity simulations. We focus on Ca$_2$Fe$_3$O$_8$, CaCo$_2$O$_4$, and CaMn$_2$O$_4$, and use density functional theory (DFT) to determine their electronic and magnetic ground states and assess their microscopic MIT mechanisms. We further apply machine learning regression models to estimate their transition temperatures and employ synthesis prediction tools to identify likely precursors and reaction routes. This integrated approach reduces the time and effort required to identify, understand, and synthesize new MIT materials, providing a generalizable pathway for accelerating correlated quantum materials discovery.
\end{abstract}

%\keywords{}

\maketitle


\section{Introduction}

%Electronically active correlated electron materials and their properties can be challenging to predict, requiring high-fidelity, resource-intensive calculations to understand their electronic structure, followed by significant time and resource investment for their synthesis, as well as the accurate experimental measurement of their electronic properties. Further, the dependence of the predicted property on the exact electronic structure methodology - which often includes adjustable parameters - , combined with the high time and resource cost of synthesizing a novel material, make the cycle of discovery relatively slow. 

Predicting the properties of electronically active correlated electron materials is a complex task, demanding high-fidelity, resource-intensive calculations to understand their electronic structure, followed by substantial time and resource investment for synthesis and precise experimental measurement of their electronic properties. Additionally, the accuracy of the predicted properties often depends on the specific electronic structure methodology used, which typically involves adjustable parameters. This, coupled with the high costs and time required to synthesize new materials, significantly slows down the discovery process.
This challenge is particularly true for metal-insulator transition (MIT) compounds, which undergo a phase transition from a metallic state to an insulating state, or vice versa, through changes in external conditions, such as temperature, pressure, or doping \cite{shao_recent_2018,kisiel2023highresolution}, leading to a range of  applications \cite{nanoelectronicsMIT,SchullerMIT,WindowMIT, Cui2018,Hoffmann2022,VO2Neuromorphic}. Here, we focus on thermally driven MITs rather than compositionally driven materials, as the electronic state of these compounds usually can also be switched % not just by temperature, but also
with an applied electric field, mechanical stress, or light.
%a group with strong potential technological interest - however, 
There are fewer than 70 stoichiometrically distinct materials that exhibit this property, all of which we have documented in a database.\citep{AlexMLMIT}

Although these materials all show a change in electrical resistivity with temperature, their mechanisms, stoichiometry, and crystal structures are often different, and the exact mechanism is frequently debated.
The transition from the metallic phase to the insulating phase is usually triggered by a coupled change in symmetry in both the electronic phase and in crystal structure.
The interplay of these effects is often difficult to decouple experimentally and theoretically; however, recent experiments have made progress using, for example, isotope effects \citep{Javier2023} and superlattice structures. \citep{Lucia2023,NNOIvan,NNOref,ClaribelMagnetism} Theoretical methods to simulate the ground state show dependence on the computational methodology, making the prediction of new compounds difficult \cite{liang_tuning_2020, meyers_pure_2016, PhysRevLett.121.067601, Landscapes, AlexAdvancedMaterials}; complexities associated with doping and vacancies can often further complicate the picture. \citep{DefectsExperiment,CMOYongjin,Kotiuga2019} The temperature dependence of the transition poses an additional challenge, as density functional theory (DFT) is a ground state ($T=0$\,K) theory. While a zero temperature calculation may help elucidate the nature of the insulating ground state, it is difficult to predict with high accuracy whether an MIT will occur with temperature. As a result, much of the focus in the literature has been on explaining the difference between the two states for previously experimentally observed MITs. Even though computational models have so far been able to reproduce varying aspects of the MIT, and the temperature dependence of the electronic \citep{Landscapes} or structural degrees of freedom, \citep{ZungerTDependence} the approach is usually applied to compounds  with an experimentally reported transition. 
The high computational expense and required demonstration make this approach impractical for high-throughput applications. Additionally, these calculations do not offer guidance on how to synthesize a new material after its prediction.
Machine learning (ML) tools are increasingly used to understand and discover materials, to process - and find patterns - both in theoretical and in experimental data\citep{KaggleContest,PaperWithJennifer, fiedler2023b,jung2019,wang2021a,ward2017}, as well as to substitute explicit DFT calculations.\citep{chandrasekaran2019,ellis2021,fiedler2023,fiedler2023b} 
As ML predictions improve with larger training datasets, discovering materials becomes more difficult when there are few materials with the desired property, such as MIT compounds of interest here. To tackle this, we have formulated a tree-based machine learning model,\cite{AlexMLMIT} which needs fewer training examples than neural networks, along with physically informed feature choices.

Here, we use a combination of machine learning (ML) tools and detailed electronic structure calculations, to identify candidate MIT oxides and their synthesis pathways. We have previously reported a tree-based ML classifier trained on an MIT database and novel features\cite{AlexMLMIT}, and now we use it to filter compounds in the high-throughput DFT database Materials Project. We obtain 36 compounds from our filtration, from which we down-select to Ca$_2$Fe$_3$O$_8$, CaCo$_2$O$_4$, and CaMn$_2$O$_4$ for high-fidelity DFT study.  We find that the insulating state of Ca$_2$Fe$_3$O$_8$ is driven by a combination of trigonal-symmetry induced orbital order and magnetism, similar to what has been identified as important to the electronic structure of two-dimensional van der Waals dihalides and trihalides (MX$_2$ and MX$_3$), as well as in face and edge-connected octahedral perovskites \citep{Alex2D,NickTrigonalPerovskites}. 
The ground state of Ca$_2$Fe$_3$O$_8$ hosts an unusual Fe$^{4+}$ ($d^4$) configuration, which we show is stabilized by the trigonal crystal field imposed by the oxide anions. 
In contrast, CaCo$_2$O$_4$ hosts a a mechanism similar to that reported in LaCoO$_3$ \cite{Rondinelli2009, Sterbinsky2018,Kwon2014b} with a low-spin insulating state and a high-spin metallic state. 
CaMn$_2$O$_4$ is found to have a magnetically-driven MIT. 
Last, we propose synthesis precursors and possible transitions temperatures for the select compounds using ML methods. 
%using natural language processing. 
%
These predicted synthesis pathways should lower the time lag between identification of promising MIT compounds and validation of their crystal structures and physical  properties.

\section{Methods}

\begin{figure}[t]
	\centering
	\includegraphics[width=0.8\textwidth]{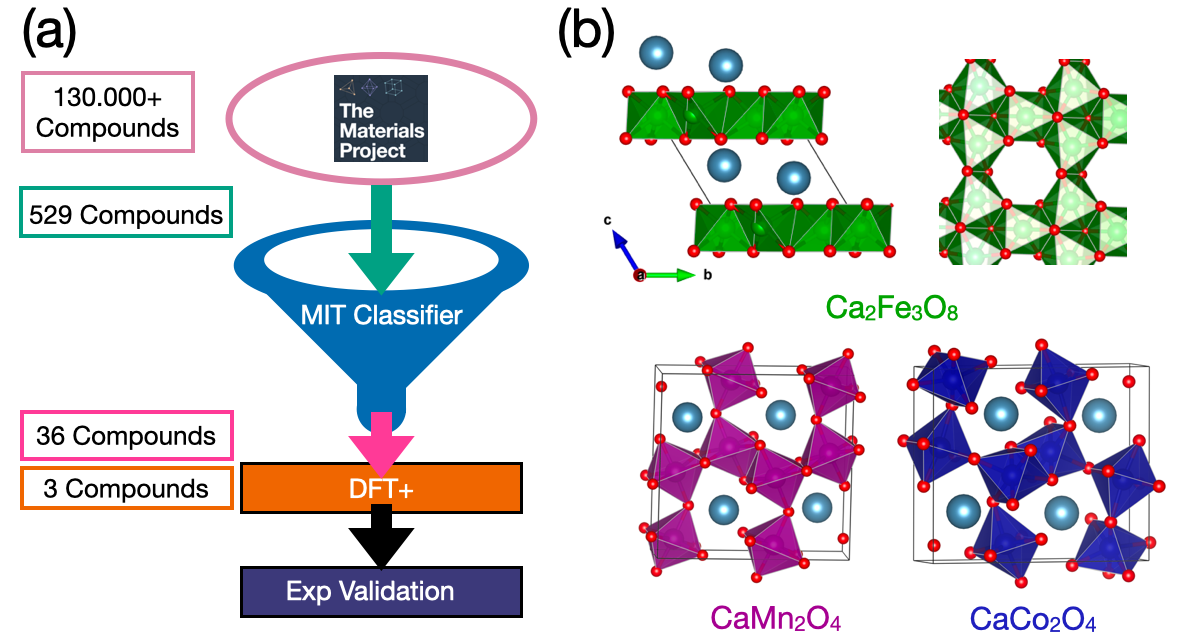}
	\caption{(a) Workflows for MIT materials identification. A curated set of compounds found in the Materials Project were classified by our ML model, which we then  performed DFT calculations on and provided synthesis procedures for experimental validation. (b) Crystal structures of the candidate MIT compounds. The top right image shows the structure of a single layer of Fe-O polyhedra, which may lead to complex spin structures.}
\label{fig:methodology}
\end{figure}

Our computational methodology includes multiple steps (\autoref{fig:methodology}). The first, which was reported in previous work,\citep{AlexMLMIT} is training a tree-based (XGboost) ML-based classifier on a wide variety of materials, consisting of all the known materials with a thermally-driven metal-insulator transition at the time of building the classifier. %  \footnote{
This classifier uses a combination of features that are known from previous work as being key to whether a material is an MIT or not (unscreened Hubbard $U$, charge transfer energy), structural features related to relevant energy scales (metal-metal, and metal-ligand distances), and other features which were found to be relevant via F1 scores by the classifier model (Global Instability Index, Average Deviation of the Covalent Radius).%} 

The second step is to use the ML-classifier to scan over a high throughput database of crystal structures, namely Materials Project\citep{matproj}, within a range of parameters where we believe that the classifier predictions are at their most reliable. The materials space scanned should, of course, include materials not in our dataset, but should not stray too far in order to be reliable. To interface our classifier with the Materials Project, we use the Pymatgen package to interface with the Materials API \cite{ong_materials_2015}. There are 3 query parameters: the chemical composition, energy above the convex hull, and the number of atoms in the unit cell. 
The chemical composition search space is constrained to A$_m$B$_n$X$_k$ ternary compounds, with elements A and B as the cations, X as the anion, and m, n, k arbitrary non-zero integers, with A=$\{$Ca, Sr, Ba, La$\}$,  B=$\{$Ti, V, Cr, Mn, Fe, Co, Ni, Cu, Zr, Nb, Mo, Tc, Ru, Rh, Pd,Ag$\}$, and X = $\{$N, S, O$\}$. In order to limit ourselves to materials that are possible to synthesize, we limit the search to materials that less than  100\,meV per atom above the convex hull.

A total of 529 structures were then pulled from the MP database. 
After passing these compounds through the MIT binary classifier, 59 structures not in the training set were predicted to exhibit a MIT; as multiple structures are polymorphs of the same compound, there are only 36 unique chemical formulas as a result of our filtration.
These results of this analysis are presented in  \autoref{tab:new_mits}, with their respective d-electron count.

\begin{table}[t]
\begin{tabular}{lcclcclc}
\toprule
Compound & \multicolumn{1}{c}{$d^n$} & & Compound & \multicolumn{1}{c}{$d^n$} & & Compound & \multicolumn{1}{c}{$d^n$} \\ 
\cmidrule{1-2}\cmidrule{4-5}\cmidrule{7-8}
CaMn$_4$S$_8$  & 1,2 & & Ca$_4$Mn$_4$O$_{11}$ & 3,4    &      & Ca$_2$Co$_3$O$_8$  & 5   \\ 
Ca$_2$Cr$_3$O$_8$ & 2&    & CaMn$_4$O$_8$   & 3,4  &        & Ca$_3$Co$_2$O$_7$  & 5   \\ 
CaCrN$_2$   & 2     & &  LaMn$_2$O$_5$   & 3,4  &        & CaCoO$_3$    & 5   \\ 
CaCr$_4$O$_8$  & 2, 3& &  BaMn$_3$O$_6$   & 3, 4 & & CaCo$_2$O$_4$   & 6   \\ 
CaNi$_2$O$_8$  & 3  & &  CaMn$_3$O$_6$   & 3, 4 & &  Ca$_2$Co$_2$O$_5$  & 6   \\ 
Ca$_2$Mn$_3$O$_8$ & 3 &   & Ca$_2$Fe$_3$O$_8$  & 4   &         & Ca$_3$Ni$_2$O$_7$  & 6   \\
Ca$_3$Cr$_3$N$_5$ & 3 &   & Ca$_2$Mn$_2$O$_5$  & 4   &         & CaNiO$_3$    & 6   \\ 
CaCr$_2$O$_4$   & 3   &         & Ca$_3$Fe$_2$O$_7$  & 4  &          & BaNi$_4$O$_8$ & 6, 5 \\ 
CaMnO$_3$    & 3   &         & CaMn$_2$O$_4$   & 4 &  & CaNi$_2$O$_4$   & 7   \\ 
BaMn$_6$O$_{12}$  & 3, 4  & & CaMn$_4$N$_4$   & 4, 5   & & \\ 
BaMn$_4$O$_8$   & 3, 4   &       & BaFe$_4$O$_7$   & 5    & & \\ 
\bottomrule
\end{tabular}
\caption{Table of materials identified by the classifier as possibly displaying an MIT, as sorted by their transition metal d-orbital filling, $d^n$. }
\label{tab:new_mits}
\end{table}

Next, we curate the identified materials, labeling their current experimental status, and eliminating materials with less than one electron per d-shell (in an ionic model) for the transition metal atoms, namely CaMoN$_3$, BaCr$_2$O$_7$,  BaMo$_3$O$_{10}$, BaNb$_4$O$_{11}$, CaNb$_2$O$_6$. Finally, we perform density functional theory (DFT) calculations on the compounds where the experimental status is not known, to assess the possibility of a metal-insulator transition and its mechanism on three compounds, \cfo{}, \cco{} and \cmo.  Our DFT calculations use the Vienna Ab initio Simulation Package (VASP) \cite{PhysRevB.47.558, KRESSE199615, PhysRevB.54.11169}; unless otherwise mentioned, the SCAN meta-GGA exchange-correlation functional \cite{sun_accurate_2016} is used, which has been proven to give electronically accurate ground states on MIT and correlated insulating materials with $3d$-orbital transition metal ions \cite{PhysRevB.100.035119}. 
For the Fe and the Mn compounds, we use a non-magnetic calculation as a proxy for the metallic phase.
To predict hypothetical synthesis parameters for the identified possible MIT compounds, we use ML methods developed in \cite{karpovich2022inorganic} to predict precursors, synthesis route, and solid-state synthesis temperatures.

% \subsection{}
% \subsubsection{}

\section{Results and Discussion}

\begin{figure}[t]
	\centering
\includegraphics[width=0.8\textwidth]{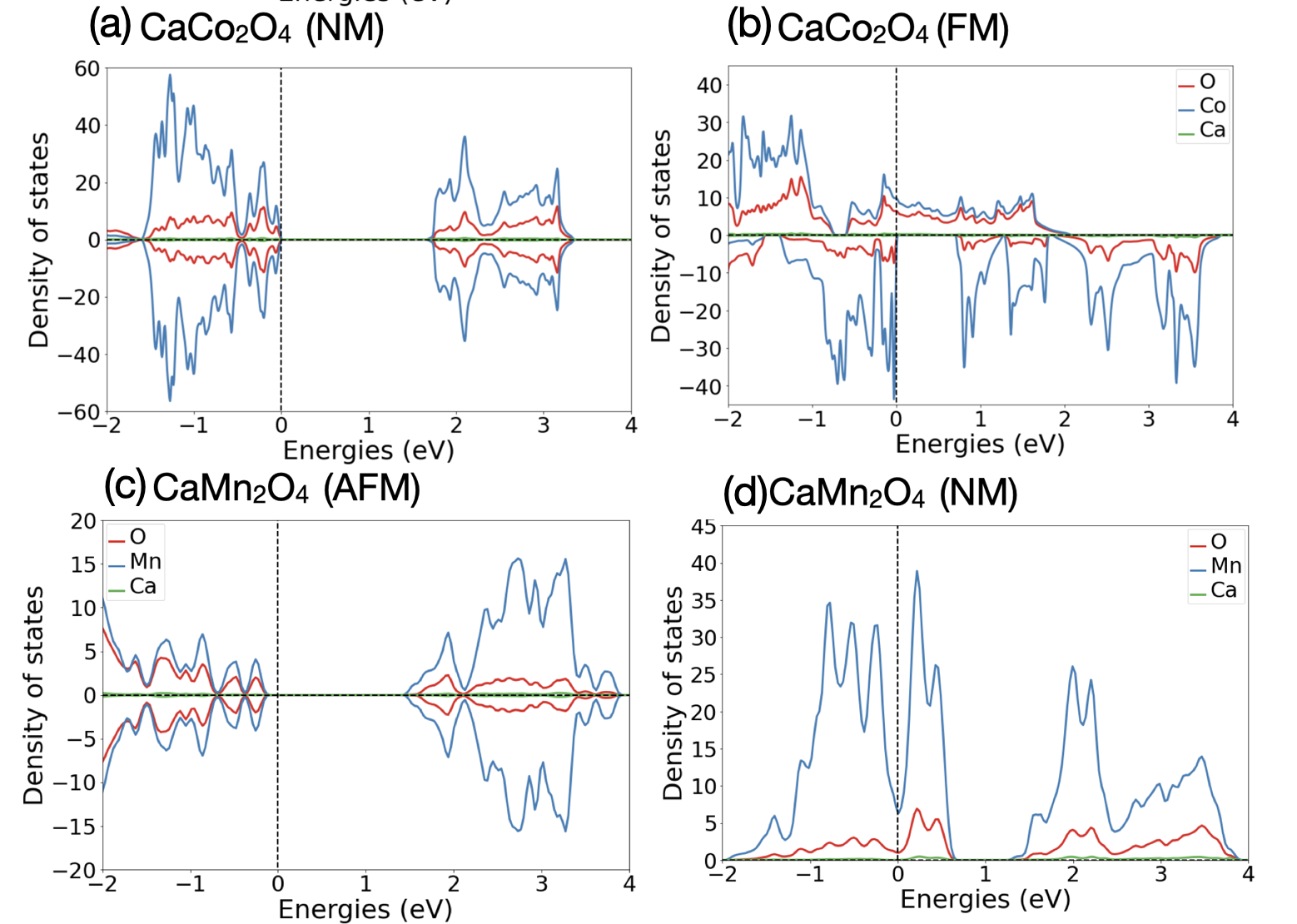}
\caption{a) and b) Density of states for CaCo$_2$O$_4$ in the non-magnetic S=0 and FM $S=1$ state, show a similar possible transition mechanism as in LaCoO$_3$, with the ground state non-magnetic, and a spin-driven transition as the temperature is raised. c) and d) Density of states for CaMn$_2$O$_4$ suggests the possibility of a magnetism-driven transition. 
\label{fig:CMOCCOdos}}
\end{figure}

CaCo$_2$O$_4$ exhibits a network of both edge and corner sharing CoO$_6$ octahedra, with Ca atoms interspersed. Co$^{3+}$ adopts a d$^6$ electronic configuration; the Co atoms have a $C_s$ point group, corresponding to no orbital energy degeneracy. Despite the additional symmetry breaking beyond octahedral due to both corner and edge-connected octahedra, the octahedral crystal field splitting dominates, and we find that the non-magnetic $S=0$ state, with the three orbitals lowest in energies  completely filled, is the ground state. 
\autoref{fig:CMOCCOdos} shows the density of states (DOS) plots for both the low-spin ($S=0$) and intermediate-spin ($S=1$) configurations of CaCo$_2$O$_4$. 
The low-spin configuration has a band gap as dictated by the crystal field splitting (band gap of $1.67$\,eV), while the high-spin structure is metallic. These results are similar to those of the MIT in LaCoO$_3$. The Co atoms in CaCo$_2$O$_4$ and LaCoO$_3$ both have a $+3$ oxidation state, resulting in a $d^6$ electronic configuration. In LaCoO$_3$, as the temperature increases, a superposition of the intermediate and high-spin multiplet states become more populated, closing the band gap.

Antiferromagnetic states could not be stabilized in CaCo$_2$O$_4$, however we were able to stabilize $S=1$ and $S=2$ states by enforcing a total magnetization in the material and a resulting FM state: these states are metallic, and the lowest energy state is still higher in energy than the $S=0$ state, and metallic - as is the case of LaCoO$_3$. The high spin ($S=2$) state is significantly higher in energy and not shown, so we focus our discussion on the  the low-spin ground state structure. A comparison with a Hubbard $U$ correction within the DFT+U approximation is presented in the Supporting Information (SI). The band gap and energy difference between the insulating and metallic states is higher, so it is likely that the electronic transition may be sharper than in LaCoO$_3$, and is likely to happen at a higher temperature.

We propose that CaCo$_2$O$_4$ should have a similar transition from a non-magnetic insulating state at low temperatures to a metallic state at high temperatures.
In LaCoO$_3$, the ground state of the Co  $d$-electrons is the $t^6_{2g}e^0_g$ low-spin, $S=0$ configuration. The band gap is determined by the crystal field splitting between the empty $e_g$ and filled $t_{2g}$ states. As the temperature is raised, electrons are excited from the $t_{2g}$ orbitals into the $e_g$ orbitals, leading to a combination of a higher energy $t^4_{2g}e^2_g$ configuration (high-spin, $S=2$), and intermediate-spin ($t^5_{2g}e^1_g$, $S=1$). The calculations using SCAN show a higher energy difference between the insulating state and the metallic state in the new compound than in LaCoO$_3$, so we investigated how this relative energy difference may depend on the particular choice of treatment of the correlated state. To do so, we also performed calculations using the PBE functional with an additional $+U$ correction (SI), in the simplified Dudarev formalism, using the Perdew-Burker-Erzenhof (PBE) exchange correlation formalism. Our results suggest the possibility of a sharper MIT or a higher transition temperature than that of LaCoO$_3$ of $T_{MIT}=600K$, as well as the possibility that CaCo$_2$O$_4$ may always be an insulator. 

%Similar to LaCoO$_3$\citep{LaCoO3applications}, this material may have potential applications in microelectronic applications, for example in Resistive RAM memories, and in potential mm radio frequency (mm RF) devices due to its combination of sharp transition, high operating temperature, and low electric field needed to induce a transition.

CaCo$_2$O$_4$ has been synthesized in polycrystalline form \cite{shizuya_structure_2007,cco1,cco2}, but has not been studied conclusively to exclude an MIT: the polycrystalline sample did exhibit insulating behavior ($\rho$) when measured in the temperature ($T$) range of $10K$ to $380K$, leaving open the possibility of a transition at higher temperature as suggested by our calculations. The material also showed metallic-temperature-dependent large thermoelectric power. The polycrystalline nature of the sample may also contribute to extra resistivity during measurement since charge carriers can scatter off the grain boundaries. This suggests the strong possibility that new measurements on higher quality samples of this material may elaborate further on the promising electronic structure of this material.

CaMn$_2$O$_4$ is isosymmetric to the binary Mn$_5$O$_8$ compound, however its dimensionality is reduced by replacing some of the layers comprising Mn with Ca. The electronic structure as a function of magnetism is discussed in Figure \ref{fig:CMOCCOdos}(c) and (d).
CaMn$_2$O$_4$ exhibits a T$_N$ of 240K, however there are no single-crystal measurements of this material's resistivity. Suppression of magnetic order in this material may lead to a closing of its gap. The CaMn$_2$O$_4$ structure identified has been experimentally verified; the material crystallizes in the orthorhombic $Pbcm$ space group\citep{cmoexp}. Ca$^{2+}$ is bonded in an 8-coordinate geometry to O$^{2-}$ anions. Mn$^{3+}$ is bonded to six O$^{2-}$ anions, forming edge and corner-sharing MnO$_6$ octahedra, with Mn in the $C_1$ point group and as a result shows no d-orbital degeneracy. The Mn ion is in a 3+ oxidation state, a d$^4$ configuration and a high spin $S=2$ spin state, with four orbitals half filled. 

\begin{table}[h!]
\centering
\begin{tabular}{ccc}
\toprule
\textbf{Magnetic Order} & \textbf{Relative Energy (meV/f.u.)} & \textbf{Band Gap (eV)} \\
\midrule
AFM  & 17.4   & 1.43   \\
AFM2 & 0        & 1.42 \\
AFM3 & 32.3    & 1.52 \\
FM   & 50.3    & 1.07   \\
NM   & 1913   & 0      \\
\bottomrule
\end{tabular}
\caption{Total energy differences per formula unit (in meV) and band gaps (in eV) for different magnetic configurations of CaMn\textsubscript{2}O\textsubscript{4}.}
\end{table}
We have performed multiple magnetic calculations, and while our search over possible magnetic order is not exhaustive, the material is insulating even in the ferromagnetic state (which is normally the most likely to be metallic of the possible magnetic states), with a bandgap of 1.1eV. Antiferromagnetic configurations are all lower in energy than the FM state and have a band gap of 1.42-1.52eV, with the ground state at 1.42eV
We find that in the non-magnetic state, the material is metallic. 
We conclude that this material may display an MIT with magnetism playing a key role.

%\subsection{\label{sec:dft} Pseudo-2D Magnetoresistive Ca$_2$Fe$_3$O$_8$}

\begin{figure}[t]
	\centering
\includegraphics[width=0.8\textwidth]{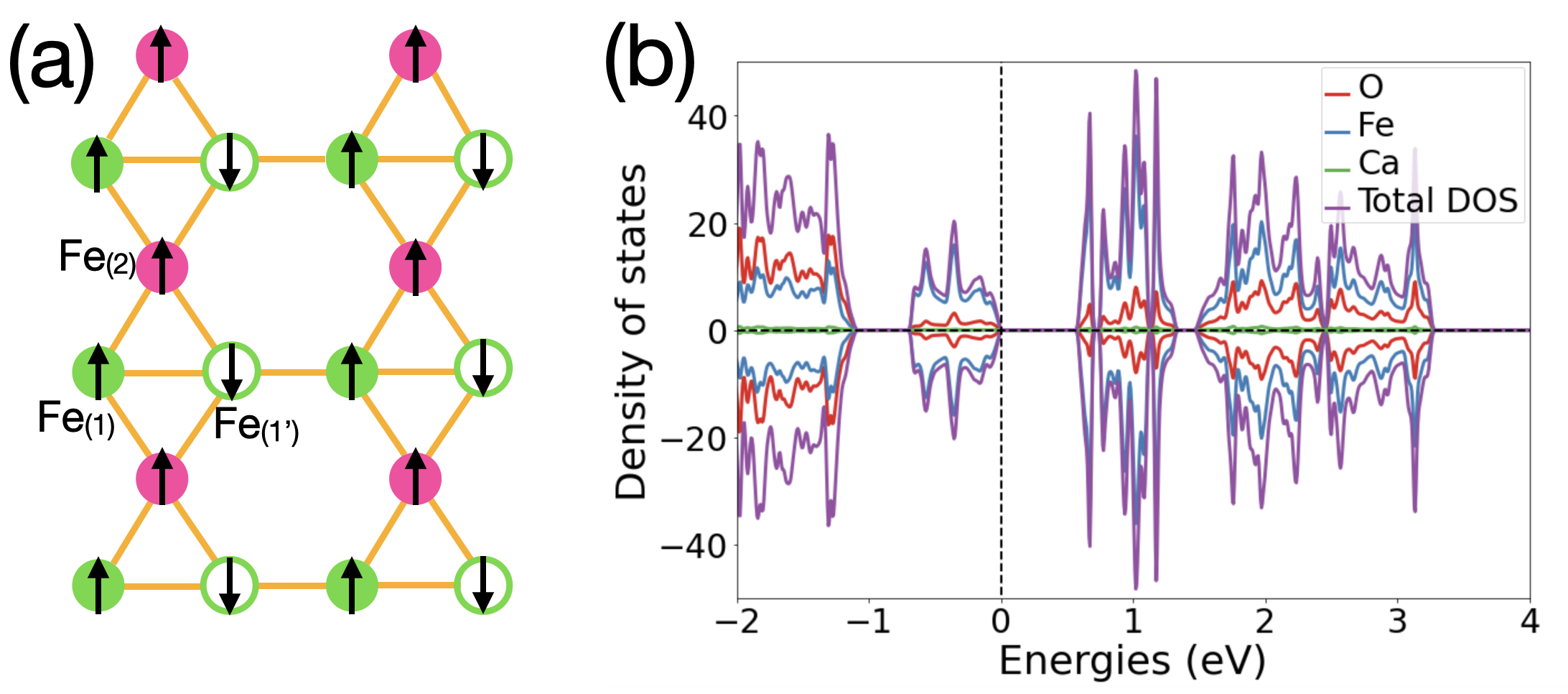}
\caption{\label{fig:cfo}(a) The magnetic order for one layer of Ca$_2$Fe$_3$O$_8$. (b) Projected density of states for the ground state of the material} 
\end{figure}

The Ca$_2$Fe$_3$O$_8$ polymorph identified by our classifier is derived from the marcasite structure and crystallizes in the hexagonal system; while there are other polymorphs on Materials Project with the same stoichiometry within 21-131 of the convex hull, neither has been experimentally confirmed. 
It is a pseudo-two-dimensional as shown in \autoref{fig:methodology}b with Ca ions filling the interstitial space between layers of FeO$_6$ octahedra. 
The 2D layers display a honeycomb lattice structure that strongly suggests the possibility of novel magnetic states, particularly as a result of spin frustration. A detailed study of non-collinear spin structures is beyond the scope of this paper, as we focus on collinear spin models . Here Fe$^{4+}$ is in a $d^4$ electronic configuration, an unusual oxidation state for oxides with octahedral crystal fields; Fe(1) has point group C$_2$ while Fe(2) C$_{2h}$; for both point groups there is no remaining d-orbital degeneracy. As we have shown in previous work \cite{Alex2D,NickTrigonalPerovskites}, edge connected octahedra with transition metal ions in a pseudo-2D material lead to an additional, trigonal splitting of the orbitals. 
The additional ligand field stabilization energy may be responsible for the relative stability of Ca$_2$Fe$_3$O$_8$ as computed by the Materials Project ($\sim$36\,meV/atom above the hull)\citep{Jain2013}. %

By doubling the unit cell to two formula units to allow for antiferromagnetic ordering, our calculations predict an anti-ferromagnetic, insulating ground state (\autoref{fig:cfo}).
The magnetic moment of each Fe site in the ground state is 2$\mu_B$, an intermediate $S=1$ spin state.
As shown in \autoref{fig:cfo}a, structurally equivalent Fe$_{(1)}$ and Fe$_{(1')}$ sites have different spin orientations, while the Fe$_{(2)}$ site has distinct symmetry (point groups $C_2$  for the (1) sites and $C_{2h}$ for (2) as previously discussed), but have the same spin state as Fe$_{(1)}$.
The magnetic order breaks the symmetry of the honeycomb, and leads to aligned spin chains within each layer, with non-zero magnetic moment per layer. The magnetic moments of alternating layers cancel out, leaving the material with net zero total magnetization. 
The point symmetry of Ca$_2$Fe$_3$O$_8$ is $C_{2h}$, a subgroup of the trigonal symmetry. As discussed in previous work, this then suggests that there is no degeneracy  in the Fe d-orbital basis in the insulating ground state. The orbital basis then consists of a fully-occupied lowest energy orbital, and two non-degenerate, half-filled, spin-polarized orbitals on each Fe atom, which is consistent with the electronic structure in \autoref{fig:cfo}b.
The first excited state is ferromagnetic and metallic. 
%the ground state found on Materials Project.
The material could display a transition to a metallic state both as a function of temperature, but also as a function of magnetic field, making it a magnetoresistive material. 
Up to the writing of this paper, this material has not been experimentally synthesized. Synthesis methods presented in this work may help guide its validation.

%\subsection{\label{sec:dft} CaCo$_2$O$_4$}

\begin{figure*}[t]
\centering
\includegraphics[width=0.95\textwidth]{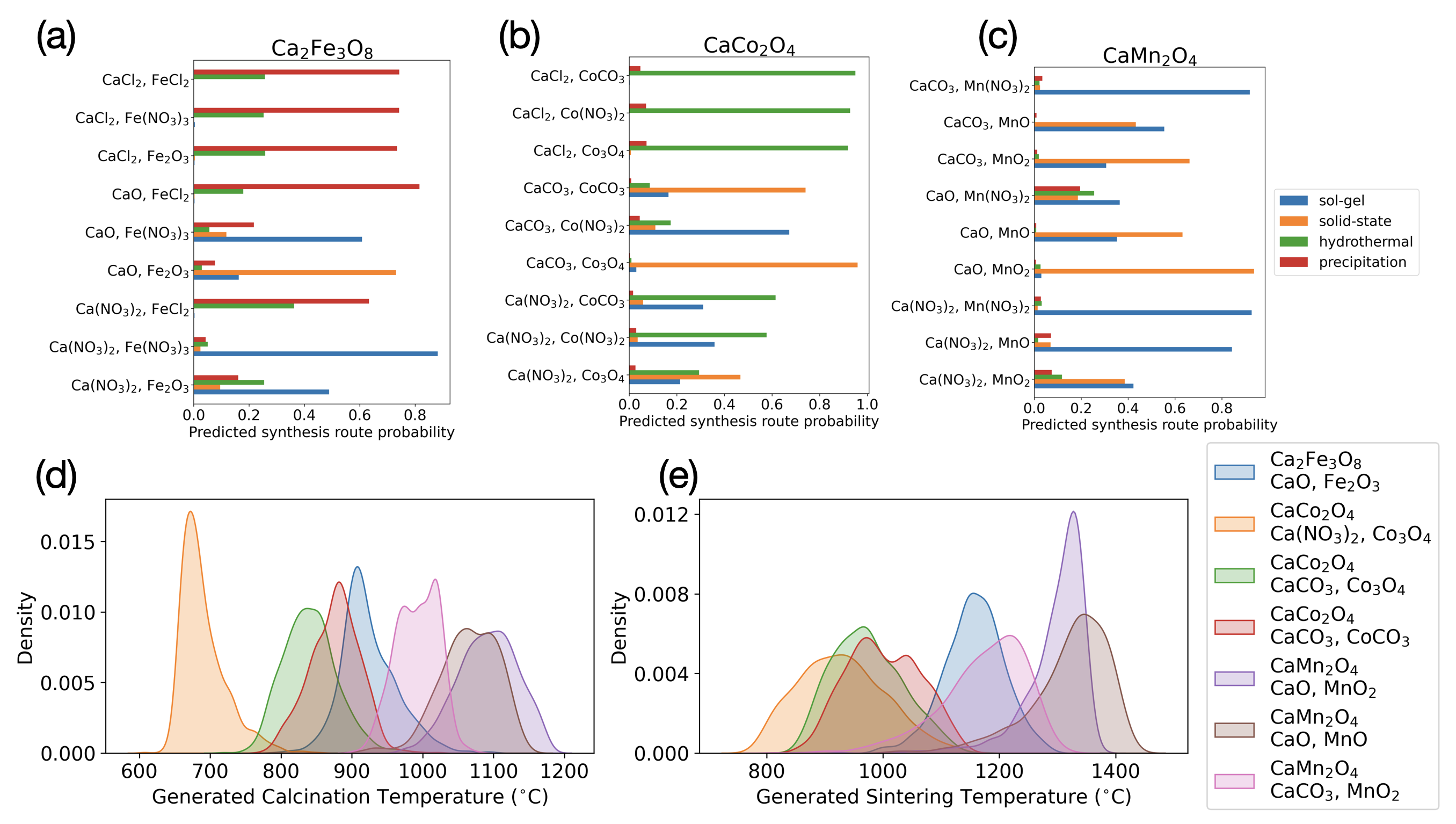}
\caption{ML-generated synthesis precursor predictions for (a) Ca$_2$Fe$_3$O$_8$, (b) CaCo$_2$O$_4$, and (c) CaMn$_2$O$_4$ and predicted (d) calcination and (e) sintering temperatures for their respective predicted solid-state synthesis routes.} 
\label{fig:synthesis}
\end{figure*}

%\subsection{\label{sec:dft} CaMn$_2$O$_4$ }

The number of known thermally driven MIT compounds is relatively small, i.e., fewer than 70, and they show a wide range of transition temperatures, even when their compositions are very similar. For instance, NdNiO$_3$ and SmNiO$_3$ differ only slightly in their Ni-O-Ni bond angles, yet their critical  temperatures differ by 200\,K. 
With these constraints, we developed a tree-based ML regression model to estimate the critical MIT temperatures.
To avoid overfitting due to this variability and limited data, we train several regression models: one using all available compounds, and two using only oxide materials.

Additional details are provided in the SI. 
Depending on the model, the transition temperatures predicted for CaMn$_2$O$_4$  and Ca$_2$Fe$_3$O$_8$  are in the 200-300\,K range, while CaCo$_2$O$_4$ is predicted to have a transition temperature in the 550-750\,K range. While it is difficult to predict the MIT temperature with high accuracy using a mean-field model, %metal-insulator transition in the Fe and Mn materials, for 
the Co compound is expected to have a higher transition temperature given the spin-state transition mechanism and the higher energy difference we find between the metallic and insulating state as compared to LaCoO$_3$. % similarity with LaCoO$_3$ 
%\subsection{\label{sec:synthesis} Synthesis Predictions}

\autoref{fig:synthesis}(a--c) shows the predicted precursors for each compound along with the corresponding probability of a solid-state, sol-gel, hydrothermal, or precipitation synthesis route. For instance, for Ca$_2$Fe$_3$O$_8$ we have 5 precipitation, 3 sol-gel, and 1 solid-state route predicted, for CaCo$_2$O$_4$ we have 5 hydrothermal, 3 solid-state, and 1 sol-gel route predicted, and for CaMn$_2$O$_4$ we have 6 sol-gel and 3 solid-state routes predicted. Plotting the predicted probabilities allows us to examine what reactions might have more than one viable option. For instance, for the synthesis of CaMn$_2$O$_4$ from Ca(NO$_3$)$_2$ and MnO$_2$ we see that sol-gel and solid-state are almost equally likely routes in terms of how promising they are to synthesize the target compound. 

From the predicted synthesis reactions for the MIT compounds, we took the ones which were predicted to be solid-state and predicted appropriate calcination and sintering temperature distributions for the reactions. \autoref{fig:synthesis} (bottom) depicts the generated calcination and sintering temperature distributions for the solid-state routes predicted for each of the MIT compounds. Lower processing conditions are desirable in many cases due to energy savings capabilities and manufacturing requirements. We see that the model predicts that for CaCo$_2$O$_4$ synthesizing the compound using Ca(NO$_3$)$_2$ and Co$_3$O$_4$ would require a much lower calcination temperature than using CaCO$_3$ and Co$_3$O$_4$ or CaCO$_3$ and CoCO$_3$. Similarly, for the synthesis of CaMn$_2$O$_4$, we see that using CaCO$_3$ and MnO$_2$ affords a lower predicted calcination temperature range than CaO and MnO$_2$ or CaO and MnO. Such predictions reveal which reactions are compatible in certain temperature ranges and thus can better inform experimental design.

\section{Conclusions}
In this work, we proposed a workflow for the computational prediction of %novel quantum materials, in this case, focusing on 
correlated materials that display a metal-insulator transition. Our current work integrated ML tools and DFT, starting with the previously built database and ML classifier, applying it to high-throughput libraries to filter materials, evaluation of possible mechanisms and synthesis of specific materials. If one or more of the materials is experimentally validated as a metal-insulator transition material, our materials discovery methodology would rapidly accelerate the discovery of new metal-insulator transition materials - and correlated electron materials more generally. Our tools are readily available online for use by other scientists.

As no reliable quantitative methods of determining suitable reaction conditions exist, our approach could help guide experimentalists with initial suggestions for synthesis precursors, route, and conditions to aid in synthesis planning for the discovery and design of new materials.
We note that we do not perform here DFT calculations to estimate which synthesis method is more likely to lead to the compounds we've predicted: these are unlikely to be reliable without an in-depth study of all the possible factors involved in the synthesis (pH, temperature, aqueous environment \cite{Walters2022}).

We note that we have also found that the classifier identifies the brownmillerite Ca$_2$Co$_2$O$_5$ as a metal-insulator transition compound: while this material may display an MIT, we find that the complex unit cell required to simulate it realistically, as well as the interplay of polymorphism and magnetic states, make it difficult to study its transition mechanism within the scope of this paper. Nonetheless, this material may display a metal-insulator transition, provided it can be grown as a single phase.

%\section{SI}

\begin{acknowledgement}
This research was supported in part through the computational resources and staff contributions provided for the Quest high performance computing facility at Northwestern University which is jointly supported by the Office of the Provost, the Office for Research, and Northwestern University Information Technology.
The information, data, or work presented herein was funded in part by the Advanced Research Projects Agency-Energy (ARPA-E), U.S.\ Department of Energy, under Award Number DE-AR0001209 (A.B.G., C.K., E.O.) and the National Science Foundation under award DMR-2324173 (P.R., J.M.R.).
This research was supported by Indiana University startup funds, and in part by the Lilly Endowment, Inc., through its support for the Indiana University Pervasive Technology Institute. 
\end{acknowledgement}

\begin{suppinfo}

The Supporting Information is available free of charge at: [\textit{\textbf{Link to be inserted}}], and relevant structures and the pipeline used to identify them can be found at: \\ https://github.com/alexandrub53/MachineLearning-MetalInsulator 

Information includes
\begin{enumerate}
    \item Comparison of CaCo$_2$O$_4$ and LaCoO$_3$ energy differences between S=0 and S=1 states for varying U
    \item Relative energies for different magnetic configurations for Ca$_2$Fe$_3$O$_8$ and CaMn$_2$O$_4$
    \item Additionally, the GitHub repository contains the scripts used to identify the materials identified in this paper, and optimized crystal structures for the three compounds studied.
\end{enumerate}

\end{suppinfo}

%\begin{tocentry}
%\includegraphics[height=1.74in]{ToC.png}
%3.25 inches by 1.75 inches (approx. 8.25 cm by 4.45 cm). 
%\end{tocentry}

\bibliography{references}

%\bibliography{references.bib,Mendeley.bib}
\end{document}